\documentclass[a4paper,two column,11pt,accepted=2021-11-02]{quantumarticle}
\pdfoutput=1
\usepackage[utf8]{inputenc}
\usepackage[english]{babel}
\usepackage[T1]{fontenc}
\usepackage{amsmath}
\usepackage[normalem]{ulem}

\usepackage[numbers,sort&compress]{natbib}

\usepackage{tikz}
\usepackage{lipsum}

\usepackage{bm}
\usepackage{physics}
\usepackage{siunitx}

\begin{document}

\title{Theory and experiment for resource-efficient joint weak-measurement}

\author{Aldo C. Martinez-Becerril}
\affiliation{Department of Physics and Centre for Research in Photonics, University of Ottawa, 25 Templeton Street, Ottawa, Ontario K1N 6N5, Canada
}
\email{amart224@uottawa.ca}

\author{Gabriel Bussi\`eres}
\affiliation{Department of Physics and Centre for Research in Photonics, University of Ottawa, 25 Templeton Street, Ottawa, Ontario K1N 6N5, Canada
}
\author{Davor Curic}
\affiliation{ Complexity Science Group, Department of Physics and Astronomy, University of Calgary, 2500 University Drive NW, Calgary, Alberta, Canada T2N 1N4 }

\author{Lambert Giner}
\affiliation{Department of Physics and Centre for Research in Photonics, University of Ottawa, 25 Templeton Street, Ottawa, Ontario K1N 6N5, Canada
}
\affiliation{ Département de Physique et d’Astronomie, Université de Moncton, 18 Ave. Antonine-Maillet, Moncton,
New Brunswick E1A 3E9, Canada }
\orcid{}
\author{Raphael A. Abrahao}
\affiliation{Department of Physics and Centre for Research in Photonics, University of Ottawa, 25 Templeton Street, Ottawa, Ontario K1N 6N5, Canada
}
\affiliation{Joint Centre for Extreme Photonics, University of Ottawa - National Research Council of Canada, 100 Sussex Drive, Ottawa, Ontario K1A 0R6, Canada}
\author{Jeff S. Lundeen}
\affiliation{Department of Physics and Centre for Research in Photonics, University of Ottawa, 25 Templeton Street, Ottawa, Ontario K1N 6N5, Canada
}
\affiliation{Joint Centre for Extreme Photonics, University of Ottawa - National Research Council of Canada, 100 Sussex Drive, Ottawa, Ontario K1A 0R6, Canada}
\homepage{http://www.photonicquantum.info/}

\maketitle

\begin{abstract}
 Incompatible observables underlie pillars of quantum physics such as contextuality and entanglement. The Heisenberg uncertainty principle is a fundamental limitation on the measurement of the product of incompatible observables, a `joint' measurement. However, recently a method using weak measurement has experimentally demonstrated joint measurement. This method [Lundeen, J. S., and Bamber, C. \href{https://doi.org/10.1103/PhysRevLett.108.070402}{ Phys. Rev. Lett. \textbf{108}, 070402, 2012}] delivers the standard expectation value of the product of observables, even if they are incompatible. A drawback of this method is that it requires coupling each observable to a distinct degree of freedom (DOF), i.e., a disjoint Hilbert space. Typically, this `read-out' system is an unused internal DOF of the measured particle. Unfortunately, one quickly runs out of internal DOFs, which limits the number of observables and types of measurements one can make. To address this limitation, we propose and experimentally demonstrate a technique to perform a joint weak-measurement of two incompatible observables using only one DOF as a read-out system. We apply our scheme to directly measure the density matrix of photon polarization states.
\end{abstract}
\section{Introduction}\label{sec:level1}
Modern quantum measurement techniques have pushed forward our understanding and ability to manipulate quantum particles. Often, fundamental and practical measurements involve the product of two or more observables of a quantum system. In particular, correlations of incompatible or non-commuting observables $\bm{A}$ and $\bm{B}$, defined by $ [ \bm{A}, \bm{B} ] \equiv \bm{A}\bm{B} -\bm{B}\bm{A} \neq 0$, are central to our understanding of entanglement \cite{einstein1935can,horodecki2009quantum} and the Heisenberg uncertainty principle. A `joint' measurement of $\bm{A}$ and $\bm{B}$ refers to the measurement process which outputs the expectation value of the product of the two observables $\expval{\bm{B}\bm{A}} =  \trace{( \bm{B}\bm{A}\bm{\rho} )}$, where $\bm{\rho}$ is the density matrix of the system. The standard procedure to perform a joint measurement would be to measure observable $\bm{A}$, then measure $\bm{B}$. This fails for incompatible observables since the first measurement collapses the state of a particle into an eigenstate of $\bm{A}$, erasing the information about $\bm{B}$ and randomizing its value. 

In contrast, weak measurement decreases the disturbance caused by the measurement process and thereby mostly preserves the quantum state of the system, thus allowing one to obtain correlations between any chosen set of general observables, including incompatible ones \cite{aharonov1990properties, resch2004extracting,lundeen2005practical,thekkadath2016direct, piacentini2016measuring, ochoa2018simultaneous, kim2018direct, chen2019experimental}. To perform such a measurement, the observable is weakly coupled to a separate read-out system (the `pointer') that indicates the average result of the measurement. Even though this approach refers to an individual system, weak measurement requires repeating the measurement on identically prepared systems, and averaging. This compensates for the little information that is extracted in a single trial. Weak measurement is  a type of non-destructive quantum measurement that minimizes disturbance of the measured system \cite{aharonov1988result}. As shown in Ref.~\cite{thekkadath2018determining}, as one decreases the disturbance caused by the measurement process, one also decreases the `predictability' of the measurement. Weak measurement has a broad range of applications from amplifying tiny signals \cite{hosten2008observation,starling2009optimizing, brunner2010measuring} to fundamental studies on the meaning of a quantum state \cite{lundeen2011direct,dressel2014colloquium}. Particularly relevant to this paper are Refs.~\cite{resch2004extracting,lundeen2005practical,mitchison2007sequential,lundeen2012procedure}, which showed that if two observables are \emph{weakly} measured, the average measurement outcome is simply the expectation value of the product of those two observables, $\expval{\bm{B}\bm{A}}$. Remarkably, this holds even if $\bm{A}$ and $\bm{B}$ are incompatible, which would make $\bm{B}\bm{A}$ non-Hermitian and, nominally, unobservable.

More recently the weak measurement formalism was expanded to deal with composite systems and performing a measurement of the product of two or more observables, known as a joint weak-measurement. Joint weak-measurement has proven to be useful, for example, in experimental realizations of the Cheshire cat \cite{aharonov2013quantum,denkmayr2014observation}, the Hardy's paradox \cite{lundeen2009experimentalHardy, yokota2009direct}, the study of quantum dynamics, and to give insight into the role of time ordering in the quantum domain \cite{dressel2017arrow,curic2018experimental}. The ability to jointly measure incompatible observables has also shown to have many applications in the field of quantum metrology \cite{ hofmann2011uncertainty,rozema2012violation, pang2014entanglement, jordan2014technical, harris2017weak, PhysRevApplied.13.034023}. Joint weak-measurement of multiple observables enables sequentially probing a quantum system for characterizing its quantum evolution \cite{kim2018direct,PhysRevA.97.042105}. Another example is the test of the Leggett-Garg inequalities for sequential measurements of multiple observables in a single system \cite{PhysRevA.80.034102}.    

Known methods for the realization of a joint weak-measurement are resource-intensive. Specifically, they require either interactions that involve three or more particles or a separate read-out system for each observable. With a few exceptions \cite{lundeen2009experimentalHardy}, due to the absence of two-particle interactions, even single-observable weak measurement resorts to a strategy of using internal degrees of freedom (DOF) as the read-out systems \cite{resch2004extracting,lundeen2005practical,mitchison2007sequential,lundeen2012procedure}. For example, one can measure the polarization of a photon by using its position DOF as a read-out \cite{ritchie1991realization}. For a joint measurement, this strategy is particularly limiting given that quantum particles have a limited number of DOF. For instance, for a photon there are just four DOF: polarization, and a three-dimensional wavevector (which, in turn, incorporates frequency-time and transverse position-momentum). Due to this limitation, joint weak-measurement experiments have never progressed beyond the product of two observables  \cite{piacentini2016measuring, thekkadath2018determining}. To overcome this constraint, the present work theoretically introduces and experimentally demonstrates a technique to perform a joint weak-measurement of multiple observables using a single DOF as the read-out system.

We implement our technique to directly measure quantum states. This is a type of quantum state estimation where the state is fully determined by the shift of the pointer.  Quantum state estimation has become an invaluable tool in the subject of quantum information, which requires verification of the quality (i.e., `fidelity') of resource quantum states. The experimental demonstration of the direct measurement of the wave function opened up new research lines in quantum state estimation. The directness of the method means that one can obtain the complex amplitudes of a quantum state, in any chosen basis \cite{lundeen2011direct}. An important aspect of direct state estimation is that no optimization or complicated inversion is involved. Solving such problems is one of the key goals of current research in quantum state estimation \cite{Glancy_2012, bolduc2017projected, PhysRevResearch.2.042002}. Further work demonstrated how to estimate a general quantum state by directly measuring the density matrix \cite{lundeen2012procedure,thekkadath2016direct}, or by  directly measuring phase-space quasiprobability distributions of states, such as the Dirac distribution \cite{salvail2013full,bamber2014observing}. Similarly, we apply our single-pointer joint weak-measurement method to directly determine any chosen element of the density matrix. Specifically, we obtain the density matrix of photon polarization states using a single pointer for the two requisite observables.

The rest of the paper is organized as follows. We start by describing weak measurement in terms of raising and lowering operators. Then, we outline the theory of our technique to perform a joint weak-measurement and introduce an important ingredient, the fractional Fourier transform. Next, we present the experimental demonstration of our technique and an application to quantum state estimation. Finally, we summarize our work and point out some future possible directions. An overview of the Fractional Fourier Transform (FrFT) is given in Appendix A.
\section{\label{sec:weakmeasurement}Weak measurement}
In this section, we introduce a theoretical model for quantum measurement, von Neumann's model, that is typically used to describe weak measurement. The model involves a measured quantum system $\mathcal{S}$ and a pointer system $\mathcal{P}$ \cite{wiseman2009quantum}. The latter indicates the measured value, the read-out, on a meter. A key aspect of the model is that the pointer is also quantum mechanical. Before the measurement, $\mathcal{S}$ and $\mathcal{P}$ are in an initial product state, $\ket{I}_\mathcal{S} \otimes \ket{\phi}_\mathcal{P}$, here $\otimes$ indicates a tensor product between different Hilbert spaces and the subscript is a label of the system. Both of these symbols will be omitted in the rest of the paper.  As usual, we assume that the pointer's initial spatial wave function $\phi(x)$ is a Gaussian centered at zero \cite{wiseman2009quantum}:

\begin{equation}
\begin{split}
 \phi(x) & \equiv \braket{x}{\phi}  \\
       &  = \frac{1}{(2\pi\sigma_x^2)^{\frac{1}{4}}} e^{-\frac{x^2}{4\sigma_x^2}},    
\end{split}
\label{eq:phistate}
\end{equation}
where $\sigma_x$ is the standard deviation of the position probability-distribution. 

The pointer's initial state happens to be same as the ground state of a harmonic oscillator. Thus, following Ref.~\cite{lundeen2005practical}, we define a lowering operator $\bm{a}$ as the operator that annihilates this pointer state, $\bm{a}\ket{\phi} = 0$. By this logic, from here on we label the pointer's initial state as $\ket{0} = \ket{\phi}$. As a standard lowering  operator, $\bm{a}$  can be written in terms of the position $\bm{x}$ and momentum $\bm{p}$ of the pointer as follows  $\bm{a} =  \bm{x}/(2\sigma_x) + i\bm{p} \sigma_x/\hbar$. (Note, we use the natural length-scale $\sigma_x$ of the system in place of the mass $m$ and angular frequency $\omega$ that usually appear in the harmonic oscillator: $\sigma_x = \sqrt{\hbar/(2m\omega)}$.) Associated with $\bm{a}$, there is a raising operator $\bm{a^{\dagger}}$ that fulfills $[\bm{a},\bm{a^{\dagger}}] = \bm{1}$. Similarly we can define number states $\ket{n}= \frac{(\bm{a^{\dagger}})^n}{\sqrt{n!}}\ket{0}$. Formulating the model in terms of lowering, raising operators and number states has proven fruitful in the past \cite{lundeen2005practical,mitchison2007sequential,lundeen2012procedure} and will be important for what follows.

Suppose we want to measure observable $\bm{A}$ of $\mathcal{S}$. Then, in the von Neumann model, one couples $\mathcal{S}$ to $\mathcal{P}$ by the following Hamiltonian, 
 \begin{equation}
 \begin{split}
\bm{H} & \equiv g\bm{A} \bm{p} \\
& = i \frac{g\hbar}{2\sigma_x} \bm{A}( \bm{a^{\dagger} - \bm{a}} ),      
 \end{split}
  \label{eq:hamiltonianwm}
 \end{equation}
here $g$ is a real parameter that indicates the interaction strength and we have used the usual decomposition of $\bm{p}$ in terms of $\bm{a}$ and $\bm{a^{\dagger}}$. We stress that there is no physical harmonic potential in the system and thus no quantum harmonic oscillator. We are following Ref.~\cite{lundeen2005practical} and simply using the formalism of raising and lowering operators to analyze the effect of the interaction on the pointer state.
 
We now consider the state of the total system after the unitary evolution induced by $\bm{H}$: $\bm{U}_A \ket{I}\ket{0} =  e^{-i \frac{t \bm{H}}{\hbar}} \ket{I}\ket{0} =  \sum_{n=0}^{\infty}  \frac{\gamma^n}{n!} \bm{A}^n ( \bm{a^{\dagger} - \bm{a}} )^n \ket{I}\ket{0} $. Here,  $\gamma \equiv \frac{gt}{2\sigma_x}$ is a unitless parameter that quantifies the measurement strength. In general, the evolved system is in an entangled state between $\mathcal{S}$ and $\mathcal{P}$. For a strong interaction ($\gamma \gg 1$), in each trial, a measurement of the position of the pointer will unambiguously indicate the value of $\bm{A}$ (though it is not particularly obvious in this harmonic oscillator formulation).
 
So far the model is general and independent of the measurement strength. Now we consider the weak measurement regime. A weak measurement is characterized by $\gamma \ll 1$, which allows one to approximate the evolved state as $\bm{U}_A \ket{I}\ket{0} = \ket{I}\ket{0} + \gamma \bm{A}\ket{I}\ket{1}$ to first order in $\gamma$. In the weak regime, the entanglement between the pointer and measured system is reduced, and the initial state $\ket{I}$ of the particle is largely  preserved. Following the work in \cite{aharonov1988result}, a post-selection on a final sytem state $\ket{F}$ is performed. Mathematically, this amounts to projecting onto $\bra{F}$ and renormalizing, after which the pointer's final state is $\ket{\phi'} = \ket{0} + \gamma\frac{\mel{F}{\bm{A}}{I}}{\braket{F}{I}}\ket{1} $. Thus the pointer's final state is largely left unchanged. That is, it is mostly left in  $\ket{0}$, but a small component proportional to $\gamma$, is transferred to $\ket{1}$ due to the interaction with $\mathcal{S}$.

Our goal is to identify in what manner the pointer is shifted by the interaction. To this end, we find the expectations of the position and momentum of the final pointer. These respectively appear as the real and imaginary parts of $\expval{\bm{a}} \equiv
 \mel{\phi'}{\bm{a}}{\phi'} =\frac{1}{2\sigma_x}\expval{\bm{x}} + i \frac{ \sigma_x}{\hbar}\expval{\bm{p}}$. Thus, using $\ket{\phi'}$ from just above, one finds

\begin{equation}
\begin{split}
 \expval{\bm{a}} &  = \gamma \frac{\mel{F}{\bm{A}}{I}}{\braket{F}{I}}  \\
    & \equiv \gamma \expval{A}_w.    
\end{split}
 \label{eq:weakvalue}
\end{equation}
Consequently, the pointer is shifted from having $ \expval{\bm{x}} = \expval{\bm{p}} = 0$  to indicating an average outcome $\expval{A}_w=\frac{1}{2\sigma_x\gamma}\expval{\bm{x}} + i \frac{\sigma_x}{\hbar\gamma}\expval{\bm{p}}$. This average pointer shift was introduced by Aharonov, Albert, and Vaidman in Ref.~\cite{aharonov1988result} and is called the  `weak value'. Unlike in the standard expectation value, $\ket{F}\neq\ket{I}$ and, thus, the weak value is a potentially complex quantity. In summary, the real and imaginary parts of the weak value are the average shifts of the position and momentum of the pointer, which, in turn, are given by the expectation value of the lowering operator.
\section{\label{sec:level2.1} Joint weak-measurement}
For composite systems, one is interested in the average value of the product of observables such as $\expval{\bm{B}\bm{A}}$. Universally, this involves correlations between two measurement outcomes (e.g., as in Bell's inequalities). In the von Neumann model, this corresponds to correlations between pointer distributions. This is true for both strong and weak measurements. In the latter case, the average outcome should be the joint weak-value,

\begin{equation}
 \expval{BA}_w \equiv \frac{ \mel{F }{\bm{B}\bm{A} }{ I } }{ \braket{F}{I}}.   
 \label{eq:jointweakvalue}
\end{equation}
A number of techniques have been proposed and demonstrated to observe the joint weak-value in pointer correlations. We now briefly review these techniques.

First, we review the case of compatible operators $\bm{A}$ and $\bm{B}$. These could be two different observables of a single particle or observables acting on two different particles. Ref.~\cite{resch2004extracting} proposed using a separate von Neumann interaction (i.e., Eq.~\ref{eq:hamiltonianwm}) and pointer for each observable (pointers 1 and 2).  This was simplified in \cite{lundeen2005practical}, which found that $\expval{\bm{a}_1\bm{a}_2} = \expval{BA}_w/\gamma^2$. This strategy of performing two separate weak measurements was experimentally demonstrated in \cite{lundeen2009experimentalHardy}.

A more challenging case, and the subject of this work, is the one in which $\bm{A}$ and $\bm{B}$ act on the same particle, but are incompatible e.g., two complementary observables such as position and momentum. Furthermore, the product $\bm{B}\bm{A}$ is not Hermitian, thus it is not considered a valid observable in standard quantum mechanics. For example, naively replacing $\bm{A}$ with $\bm{B}\bm{A}$ in the von Neumann Hamiltonian, Eq.~\ref{eq:hamiltonianwm}, results in non-unitary time evolution. However, in the weak regime, a measurement of $\bm{A}$  largely preserves the quantum state of the particle allowing a subsequent measurement of $\bm{B}$. The correlations between the outcomes of the two measurements give $\expval{BA}_w$. A technique along these lines was proposed in \cite{mitchison2007sequential}. As with the compatible observable case above, it used a separate von Neumann interaction and pointer for each observable (pointer 1 for $\bm{B}$ and pointer 2 for $\bm{A}$). In \cite{lundeen2012procedure}, the required correlation between the pointers was shown to be  $\expval{\bm{a}_1\bm{a}_2} = \expval{BA}_w/\gamma^2$ and experimentally demonstrated in \cite{thekkadath2016direct, piacentini2016measuring}. In summary, for both compatible and incompatible observables, the same technique works. The drawback of the technique is that it requires one pointer for each observable.

In particular, this requirement of one pointer per observable is resource-intensive. In most implementations of weak measurement, pointers are internal DOF of the measured particle. For example, in \cite{ritchie1991realization} a photon's polarization is measured by coupling it to the same photon's transverse spatial DOF. In absence of inter-particle interactions, this facilitates the use of weak measurement, but quickly uses up all available internal DOF. In turn, this limits the number of observables in the product and the number of DOF that can be used in the measured system for other quantum information tasks.  It is natural to ask: can we perform a joint weak-measurement with a single pointer? 

The main contribution of the present work is to introduce and experimentally demonstrate such a technique. 
Our technique uses a sequence of two standard von Neumann interactions, each given by Eq.~\ref{eq:hamiltonianwm}. Unlike the previous techniques, the two interactions couple the system to the same pointer. As in section \ref{sec:weakmeasurement}, the total initial state is $\ket{I}\ket{0}.$  The first interaction $\bm{U}_A$ couples the pointer to $\bm{A}$, while the second $\bm{U}_B$ couples the \emph{same pointer} to $\bm{B}$. The action of two von Neumann unitaries with equal interaction strength $\gamma$ is  $\bm{U}_B \bm{U}_A \ket{I}\ket{0} = e^{ \gamma \bm{B}( \bm{a^{\dagger} - \bm{a}} )  } e^{ \gamma \bm{A}( \bm{a^{\dagger} - \bm{a}} )  } \ket{I}\ket{0} = \sum_{m,n = 0}^{\infty}  \frac{\gamma^{n+m}}{n!m!} \bm{B}^m\bm{A}^n ( \bm{a^{\dagger} - \bm{a}} )^{n+m} \ket{I}\ket{0}  $. This is the final state of the total system after the two interactions.

Motivated by the techniques outlined above, which used correlations between two different lowering operators $\expval{\bm{a}_1\bm{a}_2}$, we will aim to find the expectation of the product of two identical lowering operators, $\expval{\bm{a}^2}$. Thus, we must expand the pointer state after the interaction to second order in the interaction strength $\gamma$. There are three second-order terms: $m = n = 1$; $m = 0$, $n =2$; and $m = 2$, $n = 0$. Along with the zero and first order terms, this gives
\begin{multline}
\bm{U}_B\bm{U}_A \ket{0}\ket{I} = \bigg( \ket{0} + \gamma ( \bm{A} + \bm{B}) \ket{1} +  \\
\frac{\gamma^2}{2}\big( 2\bm{B}\bm{A} + \bm{A}^2 + \bm{B}^2 \big) (\sqrt{2}\ket{2} - \ket{0}) + O\Big(\gamma^3 \Big) \bigg)\ket{I}. 
\label{eq:unitaryAB}
\end{multline}

Now we post-select the system on a final state $\ket{F}$. To second order in $\gamma$, the renormalized pointer's final state is 
\begin{multline} 
\ket{\phi'}  =  \frac{1}{\bra{F}\ket{I}} \Bigg( \braket{F}{I} \ket{0} + \gamma \bra{F} \bm{A} + \bm{B}  \ket{I} \ket{1} +  \\  \frac{\gamma^2}{2}\bra{F}  2\bm{B}\bm{A} + 
  \bm{A}^2+\bm{B}^2 \ket{I} \big( \sqrt{2}\ket{2} - \ket{0} \big) \Bigg).
\label{eq:finalphi}
\end{multline}
As per our aim, we now calculate the expectation value $\expval{\bm{a}^2}$ for $\ket{\phi'}$: 
\begin{equation}
\expval{\bm{a}^2} = 2\gamma^2\expval{BA}_w + 
 \gamma^2 \bigg( \expval{A^2}_{w} + \expval{B^2}_{w} \bigg).
  \label{eq:loweringsquared}
\end{equation}
This equation contains the weak value of the product observable $\expval{BA}_w$ but also other nontrivial weak values, $\expval{A^2}_w$ and $\expval{B^2}_w$. However, if we limit the two observables to be projectors, then $\bm{A}^2 = \bm{A} $ and $\bm{B}^2 = \bm{B} $. This turns the nontrivial weak values into single-observable weak values, which we can replace with $ \expval{A+B}_w=\expval{\bm{a}}/\gamma$. Using this and rearranging Eq.~\ref{eq:loweringsquared} to solve for $\expval{BA}_w$ we arrive at
\begin{equation}
\expval{BA}_w = \frac{1}{2\gamma^2} \bigg( \expval{\bm{a}^2}  - \gamma \expval{\bm{a} } \bigg).
 \label{eq:jointweakvalueresult}
\end{equation}
In this way, we have expressed the joint weak-value solely in terms of expectation values on the pointer's final state. However, an additional step is still necessary.
While the expectation value of a single lowering operator is easily measured in an experiment by measuring $\bm{x}$ and $\bm{p}$ in separate trials, powers of lowering operators cannot be measured as easily. To solve this, we express $\expval{\bm{a}^2}$ using $\bm{a} = \frac{\bm{x}}{2\sigma_x} + i \frac{ \bm{p} }{2\sigma_p}$, where we have used $\sigma_x\sigma_p = \frac{\hbar}{2}$ (which is valid since the pointer is in the minimum uncertainty state $\ket{0}$). Doing so, leads to the appearance of cross terms such as $ \bm{x}\bm{p}+\bm{p}\bm{x}$, which do not correspond to a straightforwardly physical read-out system observable.  

To overcome this problem, we can use the Hermitian observable $\bm{d}$ which is an equally weighted combination of $\bm{x}$ and $\bm{p}$: $\bm{d} = \frac{\sigma_d}{\sqrt{2}} \Big( \frac{\bm{x}}{\sigma_x} + \frac{\bm{p}}{\sigma_p} \Big).$ Here, $\sigma_x$, $\sigma_p$ and $\sigma_d$ are the standard deviations of the pointer in $x$, $p$ and $d$ spaces, respectively. The $\bm{d}$ observable naturally appears in a variety of quantum systems. In the Heisenberg picture in quantum optics, the $\bm{x}$ field quadrature rotates to $\bm{d}$ after an eighth of a period of oscillation; this is equivalent to an x-p phase-space rotation of $R\pi/2$, with $R=1/2$ where $R$ is the rotation order. Similarly, $\bm{x}$ rotates to $\bm{p}$ after a quarter period ($R=1$). Just as the Fourier Transform links $\bm{x}$ and $\bm{p}$, the fractional Fourier Transform (FrFT) was introduced to calculate the effect of a rotation order $R$ on a state in the Schr\"odinger picture \cite{namias1980fractional}. In summary, there are established practical methods to physically implement FrFTs and measure $\bm{d}$.

\begin{figure*}[htb!]
\begin{center}
 \includegraphics[width= 7 in]{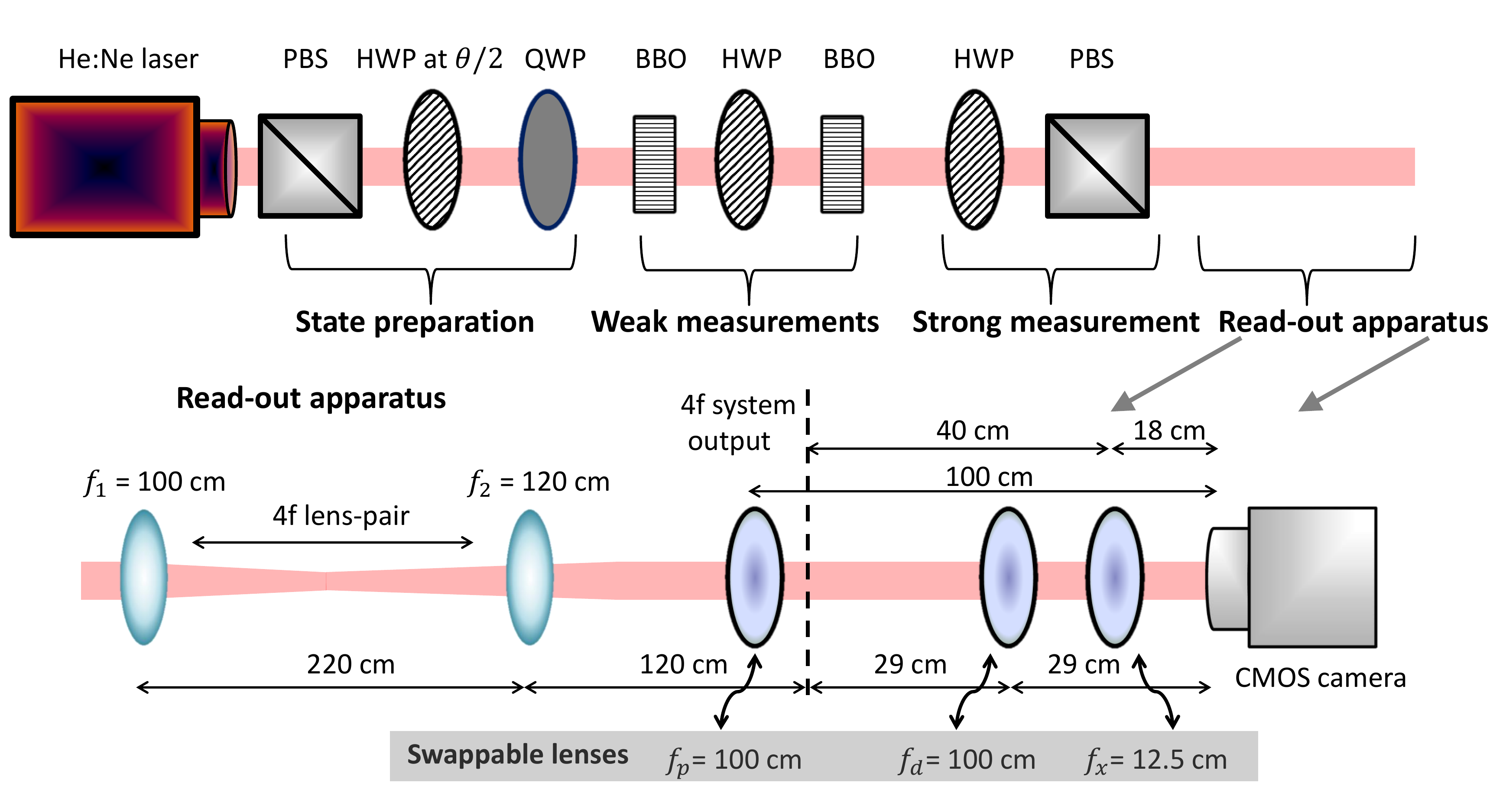}
\caption{\label{fig:setup} Experimental setup for performing a joint weak-measurement of a photon's polarization state using a single pointer, the photon's transverse $x$ position. We work with three sets of pure polarization states $\ket{\psi_1} = \cos\theta \ket{H} + \sin\theta \ket{V} $, $\ket{\psi_2} = \cos\theta \ket{H} + i\sin\theta \ket{V} $ and $\ket{\psi_3} = \frac{1}{\sqrt{2}} \big(  \ket{H} - ie^{2i\theta} \ket{V} \big)$. $\mathbf{State \ preparation:}$ To produce such states, we use a polarizing beamsplitter (PBS), a half-wave plate (HWP) set at $\theta/2$ and a quarter-wave plate (QWP). The QWP is removed for preparing $\ket{\psi_1}$, and it is set at $0^{\circ}$ and $45^{\circ}$ for $\ket{\psi_2}$ and $\ket{\psi_3}$, respectively. $\mathbf{Weak \ measurements:}$ A first walk-off crystal (BBO) implements a weak measurement of $\bm{\pi}_j$ where $j$ can be $\ket{H}$ or $\ket{V}$. A HWP at $22{.}5^{\circ}$ and a second BBO effectively perform a weak measurement of $\bm{\pi}_{45^{\circ}} = \ket{45^{\circ}}\bra{45^{\circ}}$, with $\ket{45^{\circ}} = \frac{\ket{H} + \ket{V} }{ \sqrt{2}}$. Both crystals have their optical axes aligned to create walk-off along the x-axis, the read-out DOF. $\mathbf{Strong \ Measurement:}$ A final HWP and a PBS implement a strong measurement in the $ \{ \ket{H}, \ket{V} \}$ basis. $\mathbf{Read-out \ apparatus:}$ A 4f lens-pair ($f_1 = 100$ cm and $f_2 = 120$ cm) is required to obtain the probability distributions involved in Eqs.~\ref{eq:realJWV} - \ref{eq:imagJWV} i.e., to determine $\expval{BA}_w$. For $\bm{p}$, we use a Fourier transform lens of focal length $f_p = 100$ cm, for $\bm{d}$, a Fractional Fourier Transform (FrFT) lens of focal length $f_d = 100$ cm and, for $\bm{x}$, an imaging lens of focal length $f_x = 12{.}5$ cm. Each lens is set at the specified distance from a fixed CMOS camera, the obtained images are used to calculate the required expectation values as described in the text. }
\end{center}
\end{figure*} 
The reason we have introduced this new observable is that the square of $\bm{d}$ will contain the desired cross terms. Calculating $\bm{d}^2$ and solving for the cross terms we find
\begin{equation}
 \bm{x}\bm{p} + \bm{p}\bm{x} = \sigma_x\sigma_p \Bigg( 2\frac{\bm{d}^2}{\sigma_d^2} - \frac{\bm{x}^2}{\sigma_x^2} - \frac{\bm{p}^2}{\sigma_p^2} \Bigg). 
 \label{eq:xp+px}
\end{equation}
Upon substituting Eq.~\ref{eq:xp+px} in Eq.~\ref{eq:jointweakvalueresult}, we obtain an expression for the real and imaginary parts of $\expval{BA}_w$: 
\begin{equation}
\text{Re}\Big( \expval{BA}_w \Big) =   \frac{1}{8\gamma^2} \expval{ \frac{\bm{x}^2}{\sigma_x^2} -  \frac{ \bm{p}^2}{\sigma_p^2} - \frac{gt\bm{x}}{\sigma_x^2}}
\label{eq:realJWV}
 \end{equation}
and 
\begin{equation}
\text{Im}\Big( \expval{BA}_w \Big) = \frac{1}{8\gamma^2}   \expval{ 2\frac{ \bm{d}^2 }{\sigma_d^2} - \frac{\bm{x}^2}{\sigma_x^2} - \frac{ \bm{p}^2}{\sigma_p^2} - \frac{ gt }{\sigma_x} \frac{ \bm{p} }{\sigma_p} }. 	\label{eq:imagJWV}
\end{equation}
Note that every term in  Eqs.~\ref{eq:realJWV} - \ref{eq:imagJWV} is a ratio of two variables with the same units, therefore each term is unitless. For the same reason, experimental scaling factors e.g., a magnification in the $\bm{x}$ domain, cancel out. Hence, characterization of experimental scaling factors is not required  for the use of our technique.

In summary, Eqs.~\ref{eq:realJWV} - \ref{eq:imagJWV} express the full complex joint weak-value for product observable $\bm{B}\bm{A}$ in terms of Hermitian observables on the pointer's final state. As expected, the joint weak-value appears in second order powers of $\bm{x}$ and $\bm{p}$ and our new observable $\bm{d}$.  This comprises our proposed technique to weakly measure the product of incompatible observables using only a single pointer.

\section{Experiment}
In this section, we present the experimental demonstration of our proposed technique using photons. Specifically we perform a joint weak-measurement of incompatible polarization projectors. The experimental setup is shown in Fig.~\ref{fig:setup}. The measured observable will be in the photon's polarization DOF. The pointer is the photon's transverse $\bm{x}$ position with probability-distribution given by the absolute square of the wave function in Eq.~\ref{eq:phistate} with $\sigma_x = $ \SI{403}{\micro\metre}. The photon source is a He:Ne laser at $633$ nm with a power of $1{.}19$ mW. The setup can be divided into state preparation, weak measurements, strong measurement stages, and a read-out apparatus section.  
In order to test our technique, we prepare a range of  polarization states $\ket{I} = \alpha\ket{H} + \beta\ket{V}$, where $\ket{H}$ ($\ket{V}$) is the horizontal (vertical) polarization. For state preparation, we use a polarizing beam splitter (PBS) followed by a half-wave plate (HWP), set at an angle of $\theta/2$ with respect to the $\ket{H}$ polarization, and a quarter-wave plate (QWP) (see the caption in Fig.~\ref{fig:setup} for setting details).

A von Neumann measurement of polarization can be performed with a birefringent crystal (e.g., a BBO crystal) acting as a beam displacer. This optical component transversely shifts the photon by  $\Delta x=gt =$ \SI{150}{\micro\metre} if the photon is in the $\ket{H}$ polarization state and leaves it unshifted if it is in $\ket{V}$. In this way, the crystal couples the  polarization observable $\bm{A} = \ket{H}\bra{H}$ to the photon's transverse spatial position $\bm{x}$ that plays the role of the pointer. The strong measurement regime is characterized by $\Delta x$ greater than $\sigma_x$, in which the eigenstates of $\bm{A}$ are fully separated. Our experiment is performed in the weak measurement regime where $\Delta x$ is less than $\sigma_x$. 

In order to measure the product of two observables with our technique, the setup performs two weak measurements in a row. Each von Neumann interaction (i.e., Eq.~\ref{eq:hamiltonianwm}) is achieved with a separate BBO crystal. Both crystals are aligned such that they shift the transverse profile of horizontally polarized photons in the horizontal direction $x$, leaving the transverse profile in the $y$ direction unchanged. Thus, they couple to the same pointer, the $x$ DOF. The first BBO implements a measurement of $\bm{A}=\bm{\pi}_H = \ket{H}\bra{H}$. Before the second BBO, there is a HWP oriented at $22{.}5^{\circ}$. This effectively rotates the second measured observable to $\bm{B}=\bm{\pi}_{45^{\circ}} = \ket{45^{\circ}}\bra{45^{\circ}}$, with $\ket{45^{\circ}} = \frac{\ket{H} + \ket{V} }{ \sqrt{2}}$. These two measurements and their read-out constitute an experimental application of our joint weak-measurement technique that uses a single pointer. Lastly, a strong measurement of polarization observable $\bm{\pi}_j$ ($j= H$ or $V$) is performed.

In our experiment, we need the ability to measure three incompatible observables of the pointer, $\bm{x}$,  $\bm{p}$, and $\bm{d}$. This is the read-out of the result of the weak measurement. As we will explain, lens transformations will allow us to switch between these spatial observables, transforming them to a final transverse position $x'$ on a camera. We measure the probability distribution of the observables in Eqs.~\ref{eq:realJWV} - \ref{eq:imagJWV} on a monochrome 8 bit CMOS camera with a pixel width in $x'$ of \SI{2.2}{\micro\metre}. To make room for the optical lengths required for the lens transformations we add a 4f lens-pair to the imaging system ($f_1 = 100$ cm and $f_2 = 120$ cm). The 4f is positioned such that $f_1$ is 100 cm after the crystals. This ensures that the spatial wave function at the exit surface of the second crystal is recreated 120 cm after the $f_2$ lens. Our goal is to leave the camera fixed in place while different lenses are inserted in order to measure $\bm{x}$,  $\bm{p}$, and $\bm{d}$.

To measure $\bm{p}$, and $\bm{d}$ we use an optical FrFT of the spatial DOF. The special case of rotation order $R=1$ (a standard Fourier Transform), is already widely used; the transverse position $x'$ at one focal length after lens $f_p=100$ cm is proportional to  $p$ at any distance before the lens. Hence, lens $f_p$ can be placed at any distance after the 4f lens pair as long as it is $f_p$ distance from the camera. Less common is the optical spatial FrFT, which was introduced in \cite{lohmann1993image, ozaktas1995fractional, weimann2016implementation} (more details in Appendix A). At a distance $z$ after lens $f_d=100$ cm, $x'$ will be proportional to the $\bm{d}$ observable a distance $z$ before the lens. Here, $z = f_d\tan(\frac{R\pi}{4})\sin(\frac{R\pi}{2})$. For $\bm{d}$, the phase-space rotation parameter $R$ equals 1/2 making $z = 29$ cm. This $d$ lens transformation fixes the distance of the camera from the 4f lens pair. Lastly, a single lens ($f_x=12.5$ cm) placed 160 cm after $f_2$ relays the image from the 4f lens pair. This ensures that 18 cm after $f_x$, $x'$ on the camera is proportional to $x$ at the crystals. The values of $f_x$, $f_p$, and $f_d$ were chosen so that each measured $x'$ distribution spans many pixels. By switching in one lens at a time, $f_x$, $f_p$, or $f_d$, the camera effectively measures the corresponding observable.

A key experimental simplification is that we do not need to experimentally or theoretically determine the proportionality constants between $x'$ at the camera and $\bm{x}$,  $\bm{p}$, and $\bm{d}$. The imaging magnification between $x$ and $x'$ is an example of such a proportionality constant. Since they depend on the focal lengths and lens-camera distances, these constants are difficult to experimentally determine precisely. Instead, Eqs.~\ref{eq:realJWV} - \ref{eq:imagJWV}  show that each observable is divided by the width of the pointer's initial distribution in that observable, e.g. $\bm{p}/\sigma_p$. Consequently, the units cancel and all calculations can be conducted directly in terms of $x'$, i.e. camera pixel index. 
\begin{figure}[!htbp]
 \includegraphics[width = 3.4 in ]{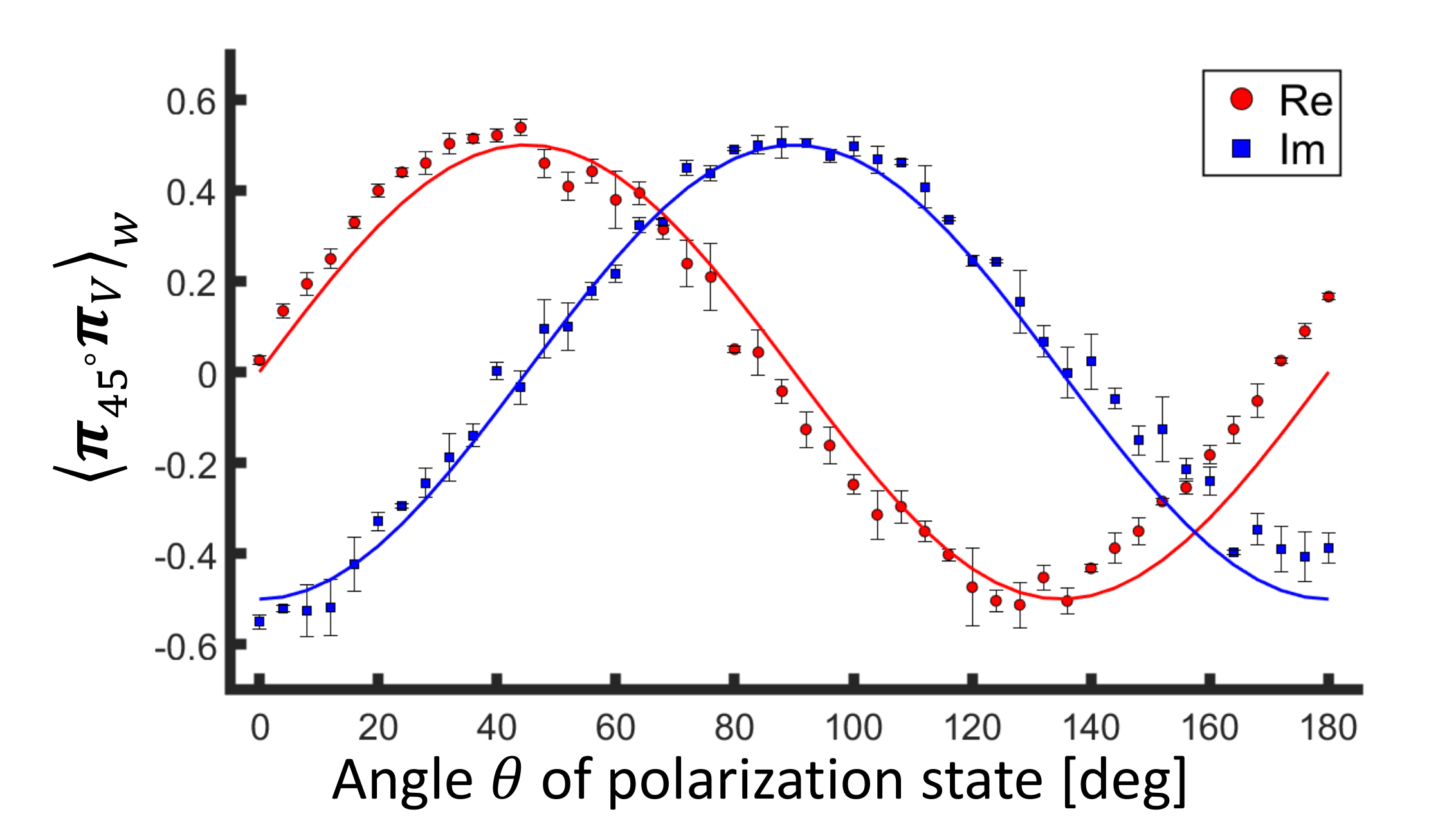}
\caption{\label{fig:weakvalue} Joint weak-value of the product of incompatible observables $\bm{\pi}_{45^{\circ}}\bm{\pi}_V$ with a  post-selection on the state $\ket{H}$. The input state is $\ket{I} = \frac{1}{\sqrt{2}} \big(  \ket{H} - ie^{2i\theta} \ket{V} \big)$ obtained by setting the preparation HWP at $\theta/2$ and the QWP at $45^{\circ}$. The real and imaginary parts of the weak value are displayed with markers, while solid lines correspond to the joint weak-value, Eq.~\ref{eq:jointweakvalue}. The error bars are calculated solely from measurement statistics and correspond to the standard deviation.
 }
\end{figure}

\begin{figure*}[t!]
\centering
 \includegraphics[width = 6 in ]{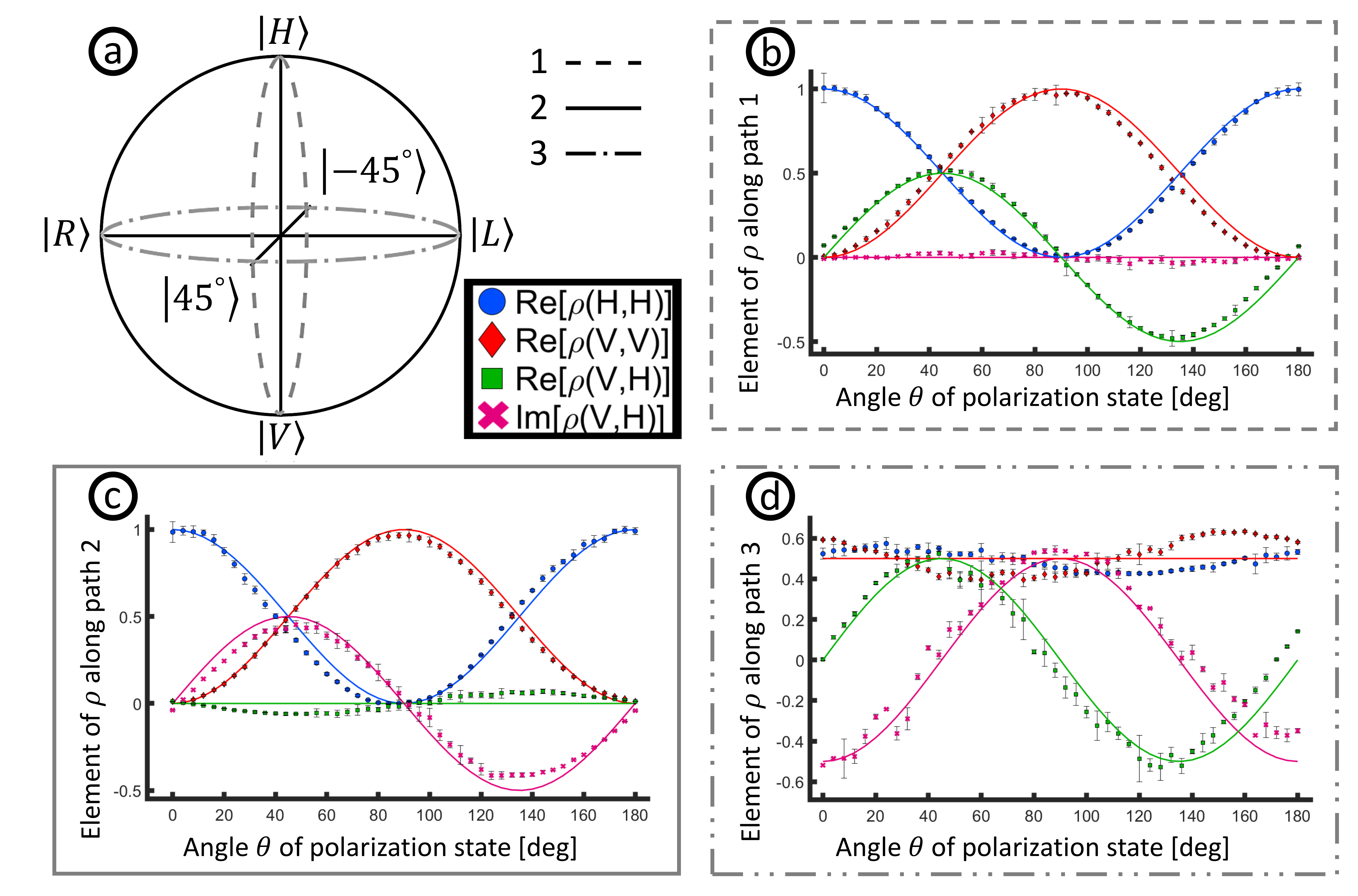}
\caption{\label{fig:results}  The polarization states used to test the method can be visualized in frame (a). These states are located on the three great circles (labeled 1, 2 and 3) in the Poincar\'e sphere passing through the states $\ket{H},\ket{V},\ket{45^{\circ}} = \frac{\ket{H} + \ket{V}}{\sqrt{2}}, \ket{-45^{\circ}} = \frac{\ket{H} - \ket{V}}{\sqrt{2}}, \ket{R} = \frac{\ket{H} - i\ket{V}}{\sqrt{2}}$ and $\ket{L} = \frac{\ket{H} + i\ket{V}}{\sqrt{2}}$. In frames (b) to (d), we show experimental elements of the density matrix $\bm{\rho}$ of the polarization states $\ket{\psi_1} = \cos\theta \ket{H} + \sin\theta \ket{V} $, $\ket{\psi_2} = \cos\theta \ket{H} + i\sin\theta \ket{V} $ and $\ket{\psi_3} = \frac{1}{\sqrt{2}} \big(  \ket{H} - ie^{2i\theta} \ket{V} \big)$, respectively. Solid lines correspond to the theory. States following path 1 corresponds to linear polarization (no QWP in the setup), states along paths 2 and 3 were obtained by setting the QWP at $0^{\circ}$ and $45^{\circ}$, respectively. Error bars are calculated using the standard deviation in the joint weak-value across the five recorded images, as previously employed in Fig.~\ref{fig:weakvalue}. }
\end{figure*}

The data acquisition consisted of taking five camera images per pointer observable (i.e., per lens configuration). A background image, taken with the laser blocked, was subtracted from each. The resulting image was integrated along the vertical direction $y'$, and normalized to the brightest image obtained in that configuration. The resulting one dimensional probability distribution $P(x')$, corresponds to the probability of detecting a photon in position $x'$ with final polarization $\ket{H}$ or $\ket{V}$. With the $f_x$ lens in place, this is effectively an $\bm{x}$ read-out. The expectation values required for the joint weak-value (Eqs.~\ref{eq:realJWV} - \ref{eq:imagJWV}) can be obtained as $\expval{\bm{x}}/\sigma_x=\expval{\bm{x}{'}}/\sigma_{x'}$, where  $\expval{\bm{x'}} =\int P(x')x' dx'$. For $\bm{p}$ and $\bm{d}$, a similar procedure is followed.

As our first demonstration of the technique, we weakly measure the non-Hermitian product observable $\bm{\pi}_{45^{\circ}}\bm{\pi}_V$ for a range of input states, $\ket{I}$. Specifically, we set the state preparation HWP at an angle of $\theta/2$ and the QWP at $45^{\circ}$ in order to produce the state: $\ket{I} = \frac{1}{\sqrt{2}} \big( \ket{H} - ie^{2i\theta} \ket{V} \big)$. We increment $\theta$ from $0^{\circ}$ to $180^{\circ}$ in steps of $4^{\circ}$. For each input state and each image, the expectation values for the joint weak-value (Eqs.~\ref{eq:realJWV} - \ref{eq:imagJWV}) were evaluated. The uncertainties were estimated by the standard deviation in the joint weak-value across the five recorded images. 

Curves for the real and imaginary parts of the joint weak-value are shown in Fig.~\ref{fig:weakvalue}. The experimental values closely follow the expected curves calculated from the nominal input state $\ket{I}$. However, they do not agree within error. These deviations are likely due to imperfections in the waveplates. These imperfections will also propagate to the alignment of the displacement axes of the BBO crystals since the waveplates are used in the alignment process. Such imperfections have been shown to be the dominant source of systematic error in similar past experiments \cite{thekkadath2016direct,Simultaneousreadout}. Nonetheless, the results demonstrate the validity of our proposed technique using a single pointer.

\section{Direct Measurement of the Quantum State}

We now move to a more sophisticated demonstration of our technique, the direct measurement of each element of the density matrix of polarization states. Such a direct measurement was introduced in \cite{lundeen2012procedure,thekkadath2016direct}. An important advantage of this direct state estimation approach is the number of measurement bases it requires. To obtain a given element of the density matrix, this method requires joint weak-measurements in two complementary bases independently of the dimension of the quantum system. This contrasts with standard quantum state tomography, which requires $O(\text{m})$ bases for an m-dimensional system. The direct estimation approach determines the density matrix of a quantum system element-by-element. One can envision a scenario where the off-diagonal elements of a system's density matrix (known as coherences) are monitored (via direct estimation) as a way to detect decoherence \cite{scholes2017using}. Unlike in Refs.~\cite{lundeen2012procedure,thekkadath2016direct}, which used two pointers for the joint weak-measurements, here we use only a single pointer (measured in three different bases) for the same task. 

A joint weak-measurement of the product $\bm{\pi}_i \bm{\pi}_{45^{\circ}} \bm{\pi}_j$, with $i,j = H$ or $V$ (with no post-selection) gives the element $\rho(i,j)$ of the density matrix. As shown in \cite{lundeen2012procedure}, the average outcome of a weak measurement without post-selection is the `weak average' (rather than the weak value), which is equal to the expectation value of the measured observable $\expval{\bm{C}}=\trace{[\bm{C}\bm{\rho}]}$. Thus, the direct measurement procedure results in a joint weak-average, $\rho(i,j) = 2\trace{[\bm{\pi}_i\bm{\pi}_{45^{\circ}} \bm{\pi}_j\bm{\rho}]}.$ Therefore, by varying the first and last projectors, the density matrix can be directly determined element-by-element.

To measure the density matrix experimentally, we changed the HWPs settings to scan over the projectors $\bm{\pi}_i$ and $\bm{\pi}_j$. As shown in \cite{lundeen2012procedure}, the final observable in the product can be measured either weakly or strongly. The last PBS implements a strong measurement. For each pair of projectors, we measure the required expectation values in Eqs.~\ref{eq:realJWV} - \ref{eq:imagJWV}. 

We test our direct measurement with three sets of pure polarization states. A general polarization state $\ket{\psi} = \alpha \ket{H} + \beta \ket{V} $ has density matrix $\bm{\rho} \equiv \ket{\psi}\bra{\psi} = |\alpha|^2\ket{H}\bra{H} + \alpha\beta^{*} \ket{H}\bra{V} + \beta\alpha^{*}\ket{V}\bra{H} + |\beta|^2\ket{V}\bra{V}.$ The states sets are $\ket{\psi_1} = \cos\theta \ket{H} + \sin\theta \ket{V} $, $\ket{\psi_2} = \cos\theta \ket{H} + i\sin\theta \ket{V} $ and $\ket{\psi_3} = \frac{1}{\sqrt{2}} \big(  \ket{H} - ie^{2i\theta} \ket{V} \big)$. These are visualized in the Poincar\'e sphere in Fig.~\ref{fig:results}a. For all cases, the first HWP varies the parameter $\theta$ scanning the interval $[0^{\circ}, 180^{\circ}]$. The QWP is removed for $\ket{\psi_1}$, and it is set at $0^{\circ}$ and $45^{\circ}$  for $\ket{\psi_2}$ and $\ket{\psi_3}$, respectively. 

Our results are shown in Fig.~\ref{fig:results}b - d. The solid lines correspond to the real and imaginary parts of the elements of the theoretical density matrix and the points are the corresponding experimental joint weak-values. The latter should be equal to the real and imaginary parts of the density matrix elements and indeed follow the expected curve for each element. As before, deviations are thought to be the result of systematic errors in the polarization optics. This direct determination of the density matrix demonstrates the utility of our technique for weak measurement applications in quantum information.
\section{Discussion and Conclusion}
Before summarizing, we discuss some special cases and extensions  of our technique. First, the special case where $\bm{B} = \bm{A}$ and the observable is general, i.e., not necessarily a projector.  In this case, the unitary evolution (Eq.~\ref{eq:unitaryAB}) of the two von Neumann interactions (i.e., Eq.~\ref{eq:hamiltonianwm}) is equivalent to a single unitary  given by $\bm{U}_A\bm{U}_A =  e^{\gamma \bm{A}(\bm{a}^{\dagger} + \bm{a})} e^{\gamma \bm{A}(\bm{a}^{\dagger} + \bm{a})} = e^{ 2\gamma \bm{A}(\bm{a}^{\dagger} + \bm{a})}$. This corresponds to a unitary of a single von Neumann interaction of the $\bm{A}$ observable with a doubled interaction strength $2\gamma$. By measuring $\bm{x}$, $\bm{p}$, and $\bm{d}$ on the pointer and using Eqs.~\ref{eq:realJWV} - \ref{eq:imagJWV}, we can then find the weak value $\expval{A^2}_w$. This behaviour can be generalized so that a single von Neumann interaction can be used to measure $\expval{A^N}_w$. To do so, one will need to measure corresponding powers of observables on the pointer, e.g., $\bm{x}^N$ as well as the \textit{N}-th power of hybrid observables such as $\bm{d}$.

In the derivation of our technique, we focused on the case that $\bm{A}$ and $\bm{B}$ are projectors. However, for general $\bm{A}$ and $\bm{B}$, measuring the product of projectors is enough to obtain the joint weak-value $\expval{BA}_w$. This can be seen if we express $\bm{A}$  in its spectral decomposition  $\bm{A} = \sum_{\alpha}\alpha \bm{\pi}_{\alpha}$. Here $\alpha$ is an eigenvalue corresponding to the eigenstate $\ket{\alpha}$, and $\bm{\pi}_{\alpha} = \ket{\alpha}\bra{\alpha}$ \cite{merzbacher1970quantum}. Analogously, for $\bm{B}$ we have $\bm{B} = \sum_{\beta}\beta\bm{\pi}_{\beta}$. Therefore, $\bm{B}\bm{A} \equiv \sum_{\alpha,\beta} \alpha\beta \bm{\pi}_{\beta}\bm{\pi}_{\alpha}$. Thus, by measuring each of the products $\bm{\pi}_{\beta}\bm{\pi}_{\alpha}$ and adding the results, the joint weak-value $\expval{BA}_w$ can be obtained. In summary, our method can be used to weakly measure the product of general incompatible observables $\bm{A}$ and $\bm{B}$.  
 
Our technique is also applicable to observables on separate quantum systems, e.g., $\bm{A}$ is measured on a first particle and $\bm{B}$ is measured on a second particle. The standard procedure for weak measurement would couple $\bm{B}\bm{A}$ to a single pointer using $\bm{H} = g\bm{B}\bm{A}\bm{p}$, which is Hermitian now. This, however, requires a three-particle interaction which is challenging. Our method uses a two-particle interaction on each system while still only using a single pointer. One would first use the standard von Neumann interaction Eq.~\ref{eq:hamiltonianwm}, to couple the pointer to $\bm{A}$, then couple the same pointer to $\bm{B}$. This double coupling can be challenging to implement. Particularly, in photons one would need an optical nonlinear effect at the single-photon level. As our technique largely preserves the initial quantum state, potential applications include demonstrating contextuality \cite{PhysRevLett.48.291, PhysRevLett.113.200401, PhysRevA.100.042116, Cimini_2020}, and tests of Leggett-Garg inequalities \cite{PhysRevLett.54.857,  PhysRevA.96.052123, PhysRevLett.106.040402, Goggin1256,PhysRevA.89.012125, white2016preserving, PhysRevA.80.034102}. Motivated by the direct measurement of entangled systems \cite{PhysRevLett.123.150402, PhysRevLett.127.030402} and entanglement witnesses \cite{PhysRevLett.92.087902, friis2019entanglement}, a potential future research direction is the use of this technique for characterizing entanglement.

A possible extension of our technique is measuring the product of \textit{m} observables of a quantum system, using a single pointer. One would perform subsequent couplings between each of the \textit{m} observables, and the pointer. The product of the \textit{m} observables will appear in the expectation value of the \textit{m} power of the lowering operator of the pointer $\expval{\bm{a}^m}$. Performing such a read-out potentially requires full-tomography on the pointer. The advantage of this approach is that only two-particle interactions are employed, and a single pointer is required for measuring the product of \textit{m} observables. 
 
As an outline of future work, the technique introduced in this paper can also be extended to general types of pointers such as spin pointers.  Indeed, previous work showed that the lowering operator formalism can be extended to spin pointers (e.g., polarization) and spin lowering operators \cite{lundeen2005practical}. To measure the product of $N$ observables, one would need to sequentially couple the $N$ observables to the spin pointer. The product of the $N$ observables will appear as a coefficient of the $N^{\text{th}}$ excited state of the pointer (just as the product $\bm{B}\bm{A}$ appears as a coefficient of the second excited state in Eq.~\ref{eq:unitaryAB}). Thus the key requirement will be that one needs $N$ spin levels (i.e., $N=2S+1$, where $S$ is the spin) to measure the product of $N$ observables. This shows that the method trades the resource of $N$ dimensions in the pointer for an   $N$-particle interaction (Eq.~\ref{eq:hamiltonianwm}) or, alternately, $N$ separate pointer systems  as in Ref.~\cite{lundeen2005practical}. Instead of encoding the measurement information in many separate pointers as in Ref.~\cite{lundeen2005practical,lundeen2009experimentalHardy,thekkadath2016direct}, we encode the information from the measurement in a single high-dimensional pointer.

As the main contribution of this paper, we theoretically derived and experimentally demonstrated a method to perform a joint weak-measurement of two incompatible observables using a single pointer. We then employed this method to directly and individually measure each element of a system's density matrix. Since  product observables are ubiquitous in quantum information processing, our technique may be useful for probing and characterizing such processors \textit{in situ}  without substantially disturbing them.  Our work optimizes the use of resources needed to perform a joint-weak measurement freeing degrees of freedom of a quantum particle for quantum information tasks. We hope our technique facilitates the use of weak measurement in complex dynamics and new studies in the quantum realm. 

\begin{acknowledgments}
This work was supported by the Canada Research Chairs (CRC) Program, the Natural Sciences and Engineering
Research Council (NSERC), the Canada Excellence Research Chairs (CERC) Program, and the Canada First
Research Excellence Fund award on Transformative Quantum Technologies. ACMB acknowledges support from Mitacs Globalink Graduate Fellowship. 
\end{acknowledgments}

\bibliographystyle{unsrtnat}
\bibliography{AldoCMB.bib}

\onecolumn\newpage
\appendix
\section{Appendix}
In this appendix we give an overview of the Fractional Fourier Transform (FrFT). Since its inception, the FrFT has been framed in different contexts: Condon described the FrFT mathematically in \cite{condon1937immersion}, while Namias framed it in quantum mechanical terms \cite{namias1980fractional}. This versatility has allowed a broad range of applications ranging from differential equations to quantum optics \cite{Alonso:11}, passing by signal processing, \cite{ozaktas2001fractional, kunche2020fractional} and telecommunications \cite{SecuringinformationFrFT}. Further details of the FrFT, and code for its numerical implementation can be found in Ref.~\cite{ozaktas2001fractional}.

The FrFT of order $R$ of a function $f(x)$ is denoted by $f_R(u) \equiv \text{FrFT}(f,R)$, and it is defined as the following transform: 

\begin{equation}
 f_R(u) = \int_{-\infty}^{\infty}K_R(u,u^{'})f(u^{'})du^{'},
\label{eq:frft_def}
\end{equation}
where 
\begin{equation}
 K_R(u,u^{'}) = \sqrt{1-i\cot{R\pi/2}} e^{i\pi\big(u^2\cot{R\pi/2} - 2uu^{'}\csc{R\pi/2} + u^{'^2}\cot{R\pi/2} \big)}, 
\end{equation}
for all real values of $R$, except for even integers. For $R = 4j$, with $j$ an integer, $K_R(u,u^{'}) = \delta(u-u^{'})$, and for $R = 4j+2$, $K_R(u,u^{'}) = \delta(u+u^{'})$. 

Particularly important are the $R=0$ and $R=1$ cases of the FrFT. These correspond, respectively, to the identity and the standard Fourier Transform (FT) operators: $f_0 = f(x)$, and $f_1 = \text{FT}(f)(p)$. A useful property of the FrFT is that it is additive in $R$, i.e., $\text{FrFT}( \text{FrFT}(f,R_2),  R_1) = \text{FrFT}(f, R_1 + R_2)$. For example, $ \text{FrFT}( \text{FrFT}(f,1), 1) = \text{FrFT}(f,2) = f(-x)$ is the parity operator. Therefore, the FrFT is a transformation that interpolates the identity and a normal FT operators.

\begin{figure}[!htbp]
\centering
 \includegraphics[width = 7.0 in ]{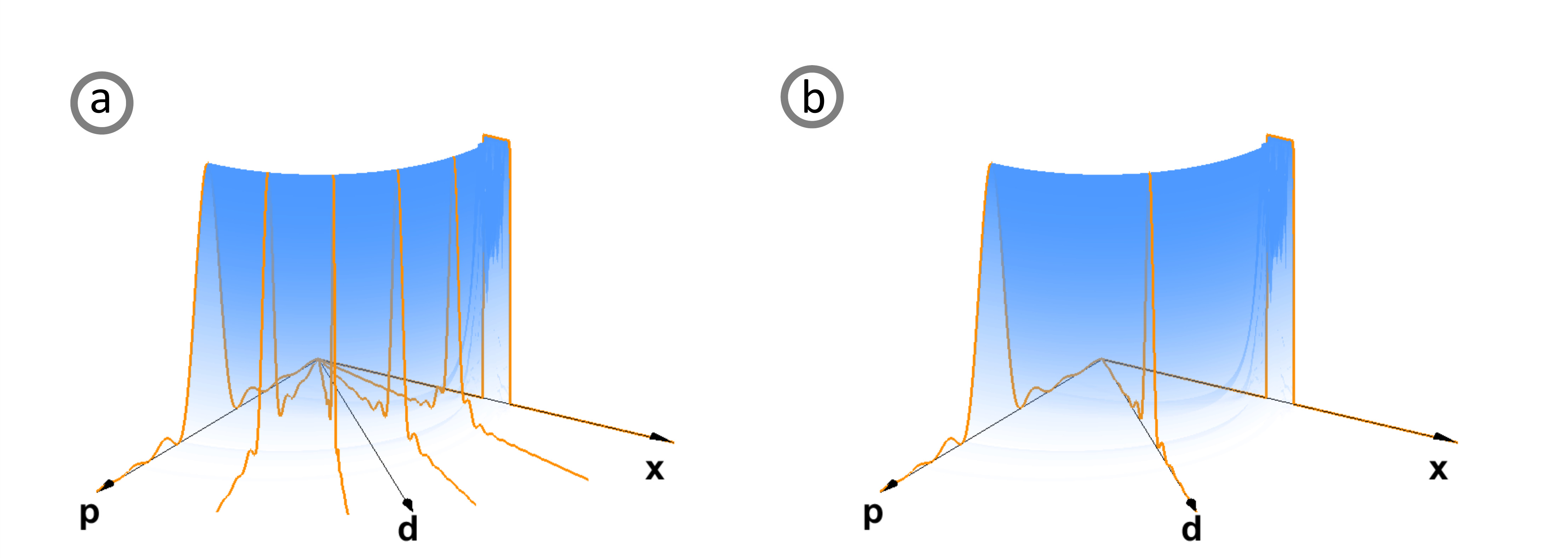}
\caption{ \label{fig:AppendixFrFT2} The Fractional Fourier Transform (FrFT) allows a continuous  transformation of a spatial function to its Fourier transform, passing through other FrFT domains. The spatial function $f(x)$ in this figure is a rectangular function. In frame (a), we are plotting $|\text{FrFT}(f,R)|^2$, for different FrFT orders $R$. Some of them are highlighted in orange, each plot is renormalized for better visualization. In frame (b), we highlight the cases of $R = 0$, $0{.}5$ and $1$, which correspond to a transformation to the $\bm{x}$, $\bm{d}$ and $\bm{p}$ domains respectively. In  the \textbf{x}-axis, we have a square function, while on the \textbf{p}-axis we observe a sinc function as we expect to be the standard Fourier Transform of the square function, in our case FrFT for $R=1$ corresponding to a $\pi/2$ rotation. The $\bm{d}$ domain is characterized by an equal weight of $\bm{x}$ and $\bm{p}$, and it shows an intensity profile between a square and a sinc functions.}
\label{FrFT_example}
\end{figure}

The FrFT is strongly rooted in position-momentum phase-space. Associated with every FrFT order $R$, there is a different quadrature $\bm{u}$ that corresponds to a different superposition of position $\bm{x}$ and momentum $\bm{p}$. For dimensionless variables, such domain can be written as $ \bm{u} = \bm{x} \cos{\b( R\pi/2 \b)} + \bm{p} \sin{\b( R\pi/2 \b)}$. The connection to phase-space is made through the Wigner distribution of a function. The action of a FrFT of order $R$ is a clockwise rotation in phase-space of the Wigner distribution by an angle $R\pi/2$. A relevant result states that a projection of the Wigner distribution  (a marginal of the Wigner distribution) on a domain at an angle $R\pi/2$, corresponds to the absolute squared of the FrFT of order $R$. In other words, performing a FrFT on a function allows one to obtain the function representation in a different FrFT domain. Fig.~\Ref{FrFT_example} illustrates an example. In the \textbf{x}-axis we have a square function, while on the \textbf{p}-axis we observe a sinc function as we expect to be the standard Fourier Transform of the square function, in our case FrFT for $R=1$ corresponding to a $\pi/2$ rotation. 

It is widely known that a FT can be performed optically by allowing free-space propagation to the far field. The FrFT completes this picture of wave propagation. Diffraction is a continuous process of FrFTs as demonstrated in \cite{pellat1994fresnel}. Optical implementations of the FrFT have been proposed utilizing lenses, graded-index media, and waveguide arrays  \cite{lohmann1993image, mendlovic1993fractional, ozaktas1993fractional, weimann2016implementation}. Now we describe the setup to perform a FrFT utilized in our implementation. 

\begin{figure}[!htbp]
\centering
 \includegraphics[width = 4.0 in ]{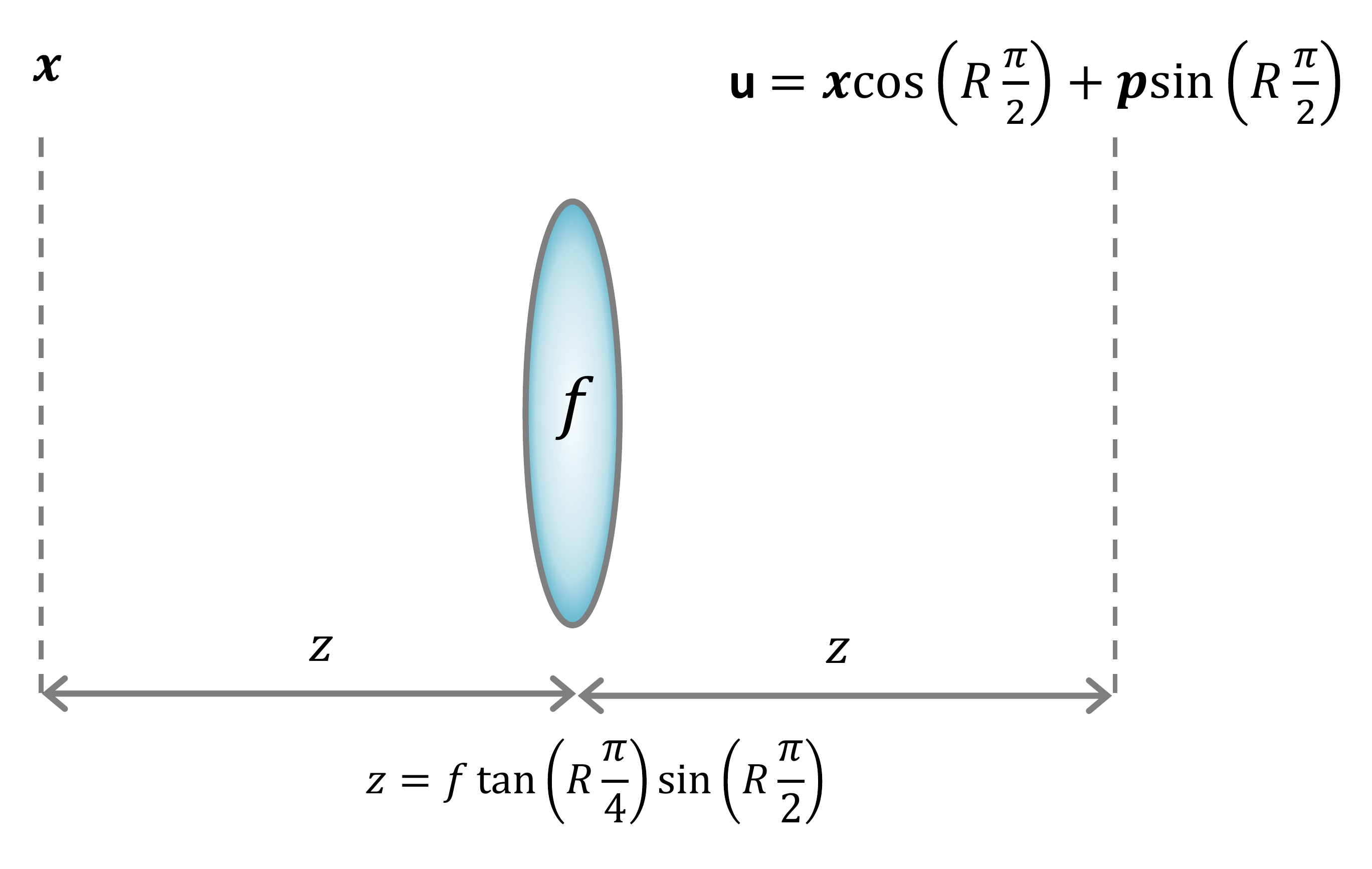}
\caption{\label{fig:sestupFrFT} The transverse position distribution of a light mode is mapped to a different FrFT domain by means of a lens and free-space propagation.}
\label{optical_FrFT}
\end{figure}

Lohman \cite{lohmann1993image} proposed a setup based on a lens that performs a FrFT of order $R$. Fig.~\ref{optical_FrFT} reproduces such setup. The transverse initial position $\bm{x}$ of an optical mode is fractional Fourier transformed by a lens of focal lenght $f$, and  free-space propagation by a distance $z = f\tan(\frac{R\pi}{4})\sin(\frac{R\pi}{2})$ before and after the lens $f$. We utilize this in our experiment for obtaining $\bm{d}$.

\end{document}